\documentclass[fleqn,usenatbib,useAMSm]{mnras}

\usepackage{newtxtext,newtxmath}

\usepackage[T1]{fontenc}

\DeclareRobustCommand{\VAN}[3]{#2}
\let\VANthebibliography\thebibliography
\def\thebibliography{\DeclareRobustCommand{\VAN}[3]{##3}\VANthebibliography}

\usepackage{graphicx}	
\usepackage{amsmath}	

\usepackage{graphicx}
\usepackage{bm}
\usepackage{natbib}
\usepackage{amsmath}
\usepackage{lineno}

\newcommand{\approptoinn}[2]{\mathrel{\vcenter{
          \offinterlineskip\halign{\hfil$##$\cr
        #1\propto\cr\noalign{\kern2pt}#1\sim\cr\noalign{\kern-2pt}}}}}

\usepackage{color}

\newcommand{\Eq}[1]{Equation~(\ref{#1})}

\newcommand{\Fig}[1]{Figure~\ref{#1}}

\graphicspath{{./fig/}{./png/}}

\title[Polar field rise rate as a precursor for solar cycle prediction]{Exploring the reliability of polar field rise rate as a precursor for an early prediction of solar cycle}

\author[Biswas et al.]{
Akash Biswas,\thanks{E-mail: akashbiswas.rs.phy20@iitbhu.ac.in}
Bidya Binay Karak,\thanks{E-mail: karak.phy@iitbhu.ac.in}
Pawan Kumar
\\
Department of Physics, Indian Institute of Technology (Banaras Hindu University), Varanasi 221005, India\\
}

\date{Accepted XXX. Received YYY; in original form ZZZ}

\pubyear{2015}

\begin{document}
\label{firstpage}
\pagerange{\pageref{firstpage}--\pageref{lastpage}}
\maketitle

\begin{abstract}
The prediction of the strength of an upcoming solar cycle has been a long-standing challenge in the field of solar physics. The inherent stochastic nature of the underlying solar dynamo makes the strength of the solar cycle vary in a wide range. Till now, the polar precursor methods and the dynamo simulations, that uses the strength of the polar field at the cycle minimum to predict the strength of the following cycle has gained reasonable consensus by providing convergence in the predictions for solar cycles 24 and 25. Recently, it has been shown that just by using the observed correlation of the polar field rise rate with the peak of the polar field at the cycle minimum and the amplitude of the following cycle, a reliable prediction can be made much earlier than the cycle minimum. In this work, we perform surface flux transport (SFT) simulations to explore the robustness of this correlation against the stochastic fluctuations of BMR tilt properties including anti-Joy and anti-Hale type anomalous BMRs, and against the variation of meridional flow speed. We find that the observed correlation is a robust feature of the solar cycles and thus it can be utilized for a reliable prediction of solar cycle much earlier than the cycle minimum---the usual landmark of the solar cycle prediction.
\end{abstract}




\section{Introduction}

The magnetic activity of the Sun determines
the space weather conditions of the heliosphere making the study of the dynamics of Sun's magnetic fields an important aspect in the field of modern Astrophysics.
Furthermore, the interaction of the Sun's magnetic fields with the magnetosphere of the Earth can cause various phenomena ranging from the occurrence of beautiful auroras in the polar regions to devastating geomagnetic storms causing malfunction of satellites, loss of communication and navigation systems, disruption in aviation near polar routes and even large scale power grid failure on the ground \citep{Gopal22}. 
With the growing technological advancement and increasing interest in space exploration over the past few decades, the study of the dynamics and impact of solar magnetism on Earth and on near-Earth space weather has gained tremendous momentum with the aim of achieving enough understanding to safeguard the space-based and ground-based assets against any adversity caused by the Sun's magnetic activities.

The strength of the Sun's magnetic field exhibits cyclic variation with a time period of about 11 years, which is known as the solar cycle.
The underlying mechanism for the solar cycle is 
believed to be dynamo action operating in the convection zone of the Sun \citep{Kar14a, Char20}.
However, the solar cycles are not identical to each other, they 
exhibit a wide range of variations in their duration and strength as generally measured by the number of sunspots that appears on the solar surface \citep{BKUW23}. 
This apparent variation in their strength makes it a very difficult but important task to predict the amplitude of an upcoming solar cycle \citep{Petrovay20,  Kar23}. 
Besides being an outstanding problem in Astrophysics, the reliable prediction of solar activity can help in the planning of upcoming space missions as well as in the study of space weather.

The Sun's magnetic field can be divided into two components, namely the poloidal and the toroidal components. In the context of the Babcock-Leighton type solar dynamo theory \citep{Ba61,Leighton69}, the dynamo operates in a cyclic fashion by producing the toroidal field through the shearing of the poloidal field by the differential rotation of the sun and the poloidal field again gets rebuilt from the toroidal field through the production and decay of the tilted bipolar magnetic regions (BMRs).

The part of the solar dynamo where the poloidal field gets built from the toroidal field experience some nonlinearities (which, at least, include nonlinear toroidal flux loss through magnetic buoyancy; \citet{BKC22}, latitude quenching; \citet{J20,Kar20} and tilt quenching; \citet{Jha20}). 
This part also involves some stochastic fluctuations \citep{KM17,BKUW23} mainly due to the inherent randomnesses in the properties of the BMRs
(presumably caused by the turbulent nature of the convection).
Observations show that the 
BMR tilt angle 
consists of significant fluctuations around 
its mean trend as given by 
Joy's law \citep{How91, Jha20} and very often BMRs are observed to have wrong tilts (negative tilts, anti-Joy) that do not obey Joy's law and BMRs (anti-Hale) that do not follow Hale polarity rule \citep{SK12, McClintock+Norton+Li14}. 
As a result of this stochastic nature, the polar field strength varies significantly from one cycle minima to another, causing
subsequent variations in the solar cycle strength \citep{Ca13, JCS14, KM18, Kit18, KMB18, Mord22}. 
It has been earlier shown that the anomalous or wrongly tilted BMRs (anti-Hale and anti-Joy type) 
can have a severe impact on the evolution of the polar field \citep[e.g,.][]{NLLPC17} and can pose a great challenge to the predictability of the solar cycles.

There has been a wide variety of approaches \citep{Petrovay20} adopted by various groups in the solar physics community for tackling the problem of predicting the amplitude of Cycles 24 and 25. 
Amongst them, the predictions based on the polar precursor methods 
\citep{Sch78, CS07, Petrovay20, Pawan21, HC19}, the dynamo 
model utilizing the polar field \citep{CCJ07,Bhowmik+Nandy}, and the surface flux transport model using observations as inputs \citep{Iijima17,Jiang18, UH18} have gained a reasonable amount of consensus due to their success in predicting the strength of Cycle 24 \citep{CCJ07, JCC07} and the convergence in their predicted strength of Cycle 25 \citep{Petrovay20, Bhowmik23, Jiang23}.
All these prediction methods are similar to some extent due to the fact that they utilize the strength of the Sun's polar magnetic fields during the solar minima as an input to predict the strength of the following cycle.

Although these methods have provided crucial knowledge about the effectiveness of the solar dynamo and the predictability of the solar cycles, a major drawback of them is that it is required to wait till the solar cycle reaches its minimum in order to reliably predict the strength of the next cycle.
On the other hand, the solar minima can be defined only after the cycle has gone past the minima phase and the next cycle has started.
In this context, it is of great importance for the solar cycle prediction community to come up with an innovative method for the reliable prediction of solar cycle strength much ahead of the solar minima. 
This can help a lot in better understanding the predictability of the solar dynamo and gain crucial lead-time in space weather awareness.

In the past, studies have cautioned about taking the polar field value much earlier than the cycle minimum for cycle prediction \citep[e.g.,][]{Sval05}.  However, \citet{Pawan21} argued that from the direct correlation between the polar precursors and the strength of the following cycle a reliable prediction can be made after 4 to 5 years of the reversal of the polar field. 
Later, \citet{KBK22} recently showed 
that, instead of taking the value of polar field strength at a particular time, if the slope of the polar field build-up (i.e., the rise rate) is considered, then 
a better correlation with the strength of the following cycle amplitude is obtained.
They argued that, the physics behind this correlation is linked with the Waldmeier effect \citep{W35} of the solar cycle and they found the prediction made using the Waldmeier effect (strong correlation between the rise rate and amplitude) matches with the prediction using the polar field rise rate.  
They used observational data from Wilcox Solar Observatory (WSO)
to calculate the average rise rate of the polar field after three years of its reversal (which occurs during the maximum of a cycle) and showed that it is highly correlated with the strength of the upcoming cycle.
Their prediction for Cycle 25 using this observed relationship converges reasonably well with other predictions made using polar precursors or dynamo models; see Table 2 of \citet{KBK22}. 
However, the reliability of the observed correlation could not be checked as the sample size available for the study is limited to only three cycles although the results from the dynamo model were consistent with observed data.

In this study, we analyze the robustness of this correlation of polar field rise rate with its amplitude and amplitude of the next sunspot cycle against the stochastic nature of BMR properties using surface flux transport (SFT) simulations. 
We introduce synthetic spatio-temporal profiles of BMRs mimicking the observed range of their properties in the SFT model to simulate the build-up of the polar field and calculate the amount of toroidal flux to be generated by the simulated polar field through differential rotation to get the strength of the following cycle. 
The main focus of the study is to understand how the variation in the distribution of tilt of the BMRs and the presence of anti-Hale and anti-Joy BMRs in varied amounts in different phases of the solar cycles impact the predictability of the following cycle using the rise rate of the polar field of the previous cycle. 
In previous years, there have been multiple studies that has shown that the meridional circulation on the solar surface changes from one cycle to another which impacts the evolution and the statistical properties of the cycles \citep[e.g.,][]{Kar10, CK12, UH14, HC17}.
Hence, we also incorporate variations in the speed of surface meridional flow to examine its impact on the aforesaid correlation.

\begin{figure*}
    
    \includegraphics[scale=0.35]{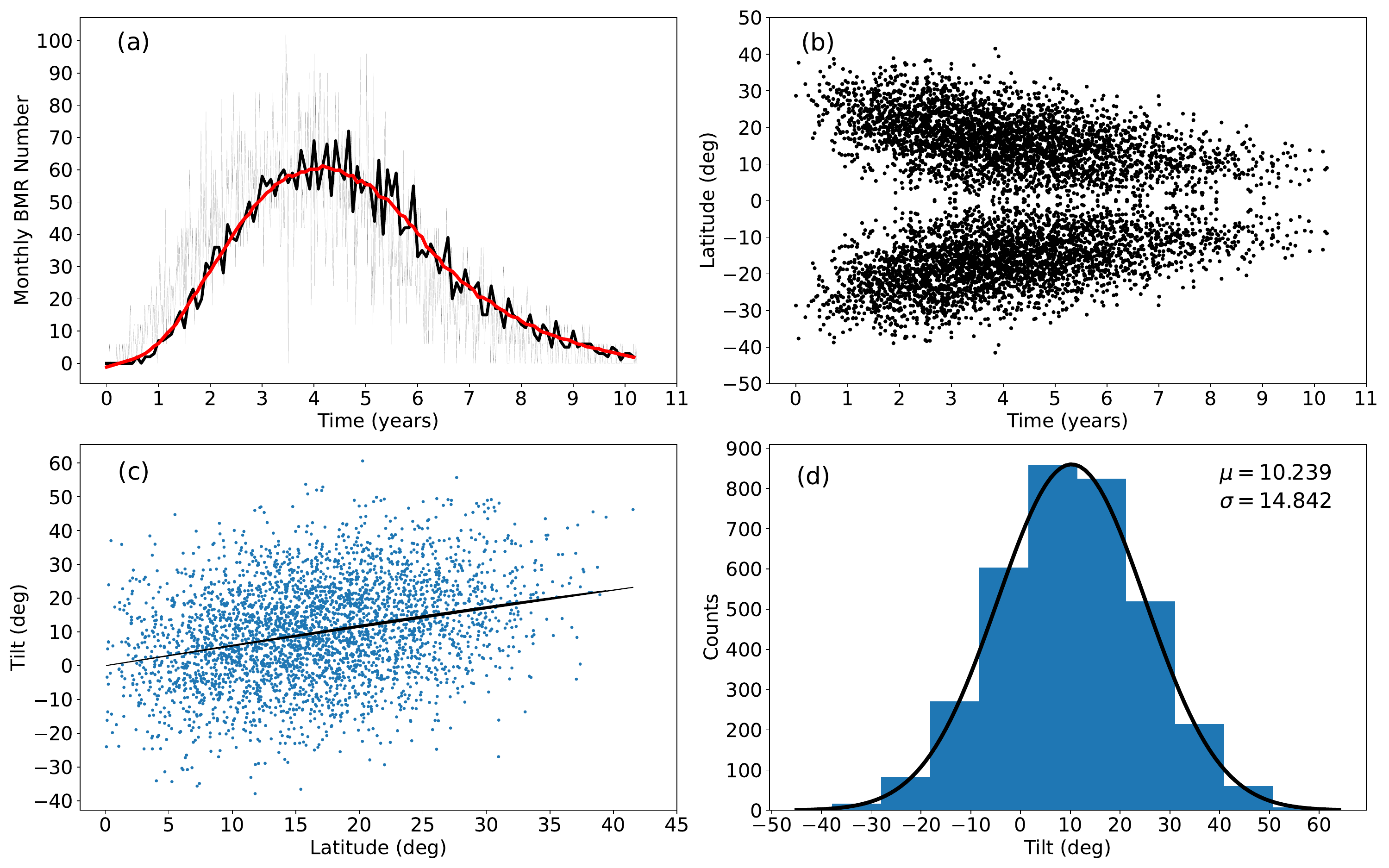}
    \caption{Profile of a synthetic solar cycle (a) and the corresponding butterfly diagram (b). Lower panels represent the properties of the typical tilt distribution used in our study. Panel (c) shows the tilts vs latitudes of BMRs (blue dots), representing the scatter of the BMR tilt around Joy's law (shown as the black solid line), and panel (d) is the distribution of the tilt and the fitted Gaussian profile with standard deviation in the similar range as found in \citet{wang15}.}
    \label{cycle_tilt}
\end{figure*}

\section{
Model and Synthetic BMR profile}
As the build-up of the poloidal field through the Babcock-Leighton mechanism takes place on the solar surface, we utilize the surface flux transport simulations for our study. Here, we briefly outline the description of the model and the spatiotemporal profile of the BMRs used as the inputs of the model.
\subsection{The Surface Flux Transport Model}
The aim of the SFT models is to capture the evolution of the radial magnetic fields on the solar surface under the influence of meridional circulation, differential rotation and horizontal diffusion \citep{WS89,Sheeley85,Bau04}. 
It captures the essence of the Babcock-Leighton framework \citep{Ba61,Leighton69} for the decay and dispersal of the tilted BMRs and the transport of the remnant diffused radial flux towards the pole due to meridional circulation which ultimately builds up the polar field. 
The governing equation at the core of the SFT model is the induction equation of Magnetohydrodynamics (MHD) which is of the following form:

\begin{equation}
 \frac{\partial \vec{B}}{\partial t} =  \vec{\nabla} \times (\vec{v} \times \vec{B} - \eta \vec{\nabla} \times \vec{B}).
 \label{eq:ind}
\end{equation}
Assuming the magnetic field is radial on the solar surface, the above equation in the spherical geometry can be written as,
\begin{eqnarray}
\frac{\partial {B_r}}{\partial t} = - \Omega(\lambda)\frac{\partial B_r}{\partial \phi} - \frac{1}{R_\odot \cos\lambda}\frac{\partial {}}{\partial \lambda} \left[ v(\lambda)B_r \cos\lambda \right]~~~~~~~~~~
\nonumber\\
+ \eta_H  \left[ \frac{1}{R_\odot ^2 \cos\lambda}\frac{\partial {}}{\partial \lambda}\left( \cos\lambda \frac{\partial {B_r}}{\partial \lambda} \right) + \frac{1}{R_\odot ^2 \cos^2\lambda}  \frac{\partial^2 {B_r}}{\partial \phi^2}  \right]
\nonumber\\
+  D(\eta_r) + S(\lambda, \phi, t).
 \label{eq:ind2}
\end{eqnarray} 
Here, $B_r$ is the surface radial field, $R_\odot$ is the solar radius, $\lambda$, and $\phi$ represent the latitude and longitude, respectively. 
The terms $\Omega(\lambda)$ and $v(\lambda)$ are the differential rotation and the meridional circulation on the solar surface which depend only on the latitude. $\eta_H$ and $\eta_r$ represent the horizontal and radial diffusivities, respectively. $D(\eta_r)$ captures the decay of the radial field due to radial diffusion and $S(\lambda,\phi,t)$ represents the source term of the radial field on the solar surface, in this case, it is the emergence of new BMRs.

For this study, we use the SFT model used in various previous studies such as, \citet{Bau04, CJSS10} etc. with similar profiles and values of the different parameters. Hence, we refrain from an elaborate discussion of the model parameters. However, we mention a few parameters of the model relevant for this work. The meridional flow on the solar surface, which has been adopted from \citet{Balle98} and is of the following form: 
\begin{equation}
v(\lambda)  = 
\begin{cases}
    v_0 \sin(2.4\lambda), &  \text{where} |\lambda|\le 75^\circ \\
    0, & \text{elsewhere}
\end{cases}
\end{equation}
where, we take $v_0 = 22$~m~s$^{-1}$ as default in the model. We use the differential rotation profile as prescribed by \citet{Snod83}: $\Omega(\lambda) = \Omega_{\rm eq} - 2.30 \sin^2\lambda - 1.62 \sin^4 \lambda$  deg~day$^{-1}$, with $\Omega_{\rm eq}$ = 13.38 deg~day$^{-1}$.

In the source term $S(\lambda,\phi,t)$, the field of each newly emerged BMR is given by $B_r = B_r^+ - B_r^-$ where,
\begin{equation}
    B_r^\pm = B_{\rm max} \left(\frac{0.4\Delta\beta}{\delta} \right)^2 \exp\left(2[1-\cos(\beta_\pm(\lambda,\phi))/\delta^2]\right)
\end{equation}
with $\beta_\pm(\lambda,\phi)$ as the heliocentric angles between $(\lambda, \phi)$ and $(\lambda_\pm, \phi_\pm)$, respectively, and $\Delta\beta$ as the separation between the two polarities, and $\delta$ as the size or area of the individual patches. The value of $B_{max}$ is taken as 374 Gauss.

\begin{figure*}    
    \includegraphics[scale=0.35]{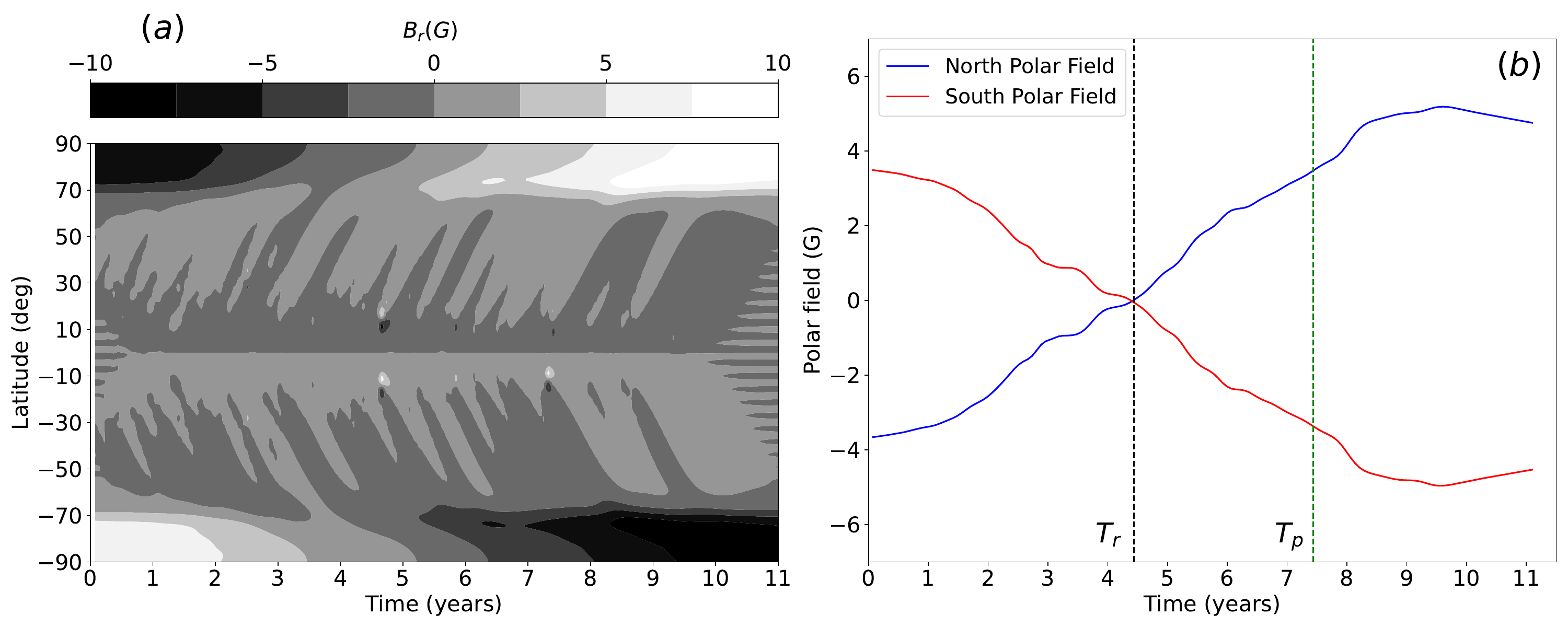}
    \caption{Evolutions of (a) the surface radial field and (b) the polar fields in the SFT simulation throughout the cycle. The black vertical line in (b) represents the reversal time ($T_r$) of the polar fields and the green vertical line represents the time up to which the polar field data is used to calculate its rise rate, i.e. this is the time ($T_p$) where prediction can be made about the strength of the 
    peak polar field 
    and about the amplitude of the next cycle.}
    \label{pol}
\end{figure*}

\subsection{The profile of the synthetic BMRs}
We perform the SFT simulations by introducing synthetic BMRs in the model as a source term with properties closely matching with the observations. To generate the synthetic profiles of the spatiotemporal evolution of BMR properties, like the variation of number of BMR emergence with time (profile of the solar cycle) or the profile of the latitudes of the BMR emergence with time (the so-called butterfly diagram) has been adopted from the analytical fittings prescribed by \cite{Jiang18} and \citet{HWR94}.

\citet{HWR94} had given a formula for obtaining the smoothed monthly number of BMR emergence in the following form 
\begin{equation}
    f(t) = \frac{a(t-t_0)^3}{\exp[(t-t_0)^2/b^2]-c},
\end{equation}
where $t$ represents time in months, $t_0$ is the time of the begining of the cycle, $a$ is the cycle amplitude. The parameter $b$ captures the Waldmeier rule of solar cycle \citep{W35, KC11}, which says stronger cycles rise quickly and is of the form, $b(a) = 27.12 + 25.15/(a \times 10^3)^{1/4}$ and $c = 0.71$.
More recently this fitting formula has been improved by \citet{Jiang18} to better fit the latest observational data of solar Cycle 12 to 24 using International Sunspot Number Version 2.0. Here we follow the formula prescribed by \citet{Jiang18} 
to obtain the monthly number of BMRs. 
Further, to obtain the daily number of BMR emergence, we divide the cycle in segments of three months and distribute the total number of emerged BMRs randomly over 90 days. This step helps in achieving the observed rapid and stochastic fluctuation in daily BMR numbers; refer to 
panel (a) of \Fig{cycle_tilt} for the profile of the solar cycle.

Once we have the variation in number of BMRs with time, we need the information of their latitude and longitudes to deposit them in the model. The longitudes of the BMRs are randomly chosen for the study. However, The observed range of the latitudinal distribution of the BMRs within a cycle depends on the strength ($S_n$) of the cycle. For the average latitude of the BMR eruptions, we utilize the same prescription given in Section~3.2 of \citet{Jiang18}. 
Here the mean of the latitude distribution of the BMRs obeys the equation: 
$\lambda_n = (26.4 - 34.2x + 16.1x^2)(\overline\lambda_n/14.6)$, where $\overline\lambda_n = (12.2 + 0.015S_n)$ and $x$ is the fraction of the solar cycle. The BMRs are randomly distributed around the mean latitude $\lambda_n$ obeying a Gaussian profile with $\sigma = (0.14 + 1.05x - 0.78x^2)\lambda_n$. In panel (b) of \Fig{cycle_tilt}, we present the butterfly diagram obtained by following the mentioned prescriptions 
and assuming the symmetry in hemisphere.

Another important parameter of the BMRs is the distribution of their area, which is crucial in determining the total radial flux content of the certain BMR. 
The BMR area ($A$ in $\mu$Hem)
has been randomly drawn from the following log-normal distribution produced from the sunspot group area data as used by previous studies such as \citet{CJSS10, Jiang11} etc.

\begin{equation}
    P(A) = \frac{1}{\sigma_a A \sqrt{2\pi}} \exp\left[-\frac{(\ln A - \mu_a)^2}{2 \sigma_a^2}\right] 
\end{equation}
where $\mu_a = 3.79$ and $\sigma_a = 0.68$.


\subsection{A note on the distribution of BMR tilt}

    
Here we discuss the most important property of the BMRs for this study, the distribution of their tilt. 
The tilt of a BMR axis with respect to the equator makes the leading polarity to emerge 
closer to the equator than the trailing polarity, which results into the trailing polarity contributing more to the build-up of the polar fields as it is situated nearer to the pole.  
Observations show
that the 
BMR tilt ($\gamma$) 
increases with the latitude of 
emergence following the equation known as Joy's law: $\gamma = \gamma_0\sin \lambda$ \citep{Hale19,WS89,How91}. 
However, 
although there is a statistical increase of the tilt with the increase in the latitude of the BMR emergence (similar to Joy's law), there exists a significant scatter in the distribution of the tilt around the value obtained from Joy's law  (\citet{How91,Fisher95,Jha20} also see Fig. 4 of \citet{Kar23}). 
This significant scatter in the tilts of the BMRs makes the contributions of the individual BMRs to the build-up of the  polar field vary dramatically. 
Very often it is observed that the tilt is even negative, which makes the leading polarity to be situated nearer to the pole and end up contributing oppositely in the build-up of the polar fields, these types of BMRs are known as the anti-Joy type BMRs \citep[e.g.,][]{JCS14,KM17}. 
In extreme cases, it is seen that the conventional longitudinal orientation \citep[according to the Hale polarity rule;][]{Hale19, SK12} of the polarities 
are flipped.
These types of BMRs are known as anti-Hale BMRs which also contribute significantly in the opposite sense to the build-up of the polar fields \citep{NLLPC17, Mord22,Pal23}. 

For the convenience of the understanding regarding the tilts associated with different types of BMRs, we divide them into four major categories as mentioned below:
\begin{itemize}
    \item Hale--Joy regions ($0^\circ<\gamma < 90^\circ$)
    \item Hale--Anti-Joy regions ($-90^\circ<\gamma < 0^\circ$)
    \item Anti-Hale--Joy regions ($-180^\circ<\gamma < -90^\circ$)
    \item Anti-Hale--Anti-Joy regions ($90^\circ<\gamma < 180^\circ$)
\end{itemize}
Here we mention that, both the Hale--Joy regions and the Anti-Hale--Anti-Joy BMRs contribute in the similar manner to the build-up of the polar fields. Hence both these type of BMRs are considered as regular BMRs, however, the latter type of regions are extremely rare in observations. On the other hand, as already discussed, the Hale--Anti-Joy (hereafter anti-Joy) and the Anti-Hale--Joy (hereafter anti-Hale) type of BMRs are considered as `anomalous' or `Rogue' BMRs \citep[for pictorial representation of these different types of BMRs see][]{munoz21,Pal23}.

Although, it is not clear whether the anti-Hale and anti-Joy type BMRs (so-called `rogue' or `anomalous' BMRs) originate through any different mechanisms than the regular BMRs or they are simply the extreme tails of the tilt distribution of the BMRs \citep{munoz21}, here we incorporate the observed statistical properties to examine their effect on the predictability of the solar cycles. 
In this work, we take the value of $\gamma_0 = 35^\circ$ following \cite{KM17} in the equation of Joy's law as mentioned above (we have taken the value of $\gamma_0$ slightly higher than the available observations, to avoid the decay of the polar field to extremely lower value in the SFT model.) 
To capture the scatter in the tilt, we impose a stochastic Gaussian noise around Joy's law, and depending on the value of the $\mu$ and $\sigma$ of the ultimate tilt distribution, a certain percentage of the total BMRs possesses a negative tilt whereas, to obtain the anti-Hale type BMRs, we randomly choose a certain percentage of total BMRs and impose appropriate tilts ($-180^\circ< \gamma < -90^\circ$) to them. The profile of tilt distribution for a certain cycle is shown in the bottom panel of the  \Fig{cycle_tilt}. We have also investigated the above-mentioned correlation under the impact of tilt scatter dependent on BMR area.
Here we mention that the convention of the tilts described here is in the perspective of the northern hemisphere, the sign of the tilt changes as we consider the southern hemisphere. 

From the information about the coordinates $(\lambda,\phi)$ and tilts of the BMRs, the coordinates of the individual poles $(\lambda_\pm, \phi_\pm)$ are calculated from the following equations: 
\begin{equation}
    \lambda_\pm = \lambda \pm \frac{\lambda}{|\lambda|}\beta_\pm \sin\gamma
\quad\mathrm{and}\quad     
    \phi_\pm = \phi \mp \frac{\lambda}{|\lambda|} \beta_\pm \cos\gamma.
\end{equation}

\subsection{Calculation of polar field and the toroidal field of the following cycle}
After introducing the synthetic BMRs in the SFT model, we run the simulations to study the evolution of the radial photospheric magnetic field. As mentioned above, the deposited BMRs decay due to diffusion and mutual flux cancellation of the opposite polarities. However, a tiny percentage of the remnant diffused radial field from the polarity situated at higher latitude (owing to the tilt of the BMR axis) end up getting advected towards the polar regions by the meridional circulation. To calculate the polar field, we first produce the magnetogram maps of the simulated photospheric fields at every 27 days intervals (Sun's rotation period at the equator). We take longitudinal averages of these maps which provide us with the latitudinal profile of the radial field. In the next step, we take the average strength of the radial field from $55^\circ$ to $90^\circ$ latitudes to obtain the strength of the polar field of each of these maps. This operation is continued for the whole cycle to get the evolution of the polar field throughout the cycle. 

For the calculation of the toroidal field to be produced in the following cycle due to the shearing of the polar field by the differential rotation of the Sun, we adopt the prescription provided by \citet{CS15}. They showed that, by applying Stoke's theorem on the mean-field form of \Eq{eq:ind} in the meridional plane of the Sun along with some simplified and reasonable assumptions we can reach to the  following equation:
\begin{equation}
    \frac{\rm d\Phi^N_{tor}}{\mathrm {d} t~~~~} = \int_{0}^{1}(\Omega - \Omega_{eq})B_rR^2_\odot \mathrm{d}(\cos\theta) - \frac{\Phi^N_{tor}}{\tau}
    \label{tor}
\end{equation}
Where, $\Phi^N_{tor}$ is the toroidal flux, $\theta (= \frac{\pi}{2}-\lambda)$ is the colatitude, and $\tau$ is a parameter representing the diffusion timescale which is taken to be four years in this work. We solve this equation taking the surface radial field $B_r$ from the SFT simulations as input to obtain the time evolution of the toroidal flux of the following cycle. 

\section{Methodology of data analysis}

In \Fig{pol} we present the evolution of the radial surface flux density $B_r$ (panel (a)) and the evolution of the hemispheric polar fields (panel (b)) for one of the cycles from our simulation. 
In the evolution of the polar field, it can be clearly seen that the strength of the polar field is strong during the beginning of the cycle (cycle minimum) and with time it decreases in strength followed by a reversal due to the fields of individual BMR that get transported to the pole by meridional flow as seen in panel (a). 
This reversal typically happens during the maxima of the cycles. 
Afterward, the polar field gets built up and reaches its peak during the next minimum due to further supply of fields from the decaying BMRs. 
For the prediction of the upcoming cycle strength from the polar precursor methods or by dynamo models, the peak of the polar field at cycle minimum is usually used.
However, the determination of the minimum of a solar cycle gets difficult due to the overlap of two consecutive cycles causing rapid fluctuation in sunspot numbers during the last few months of the declining cycle. As a result, to get the value of the polar field peak, one has to wait until the minimum of the cycle has gone past and the polar field has surpassed its peak.
In this study, following the previous work of  \citet{KBK22}, we aim to avoid this inconvenience of finding the minima of a cycle and analyze the predictability of the strength of an upcoming cycle much earlier than the minima.
We take the time of the polar field reversal ($T_r$) as the reference time in our calculations, as finding out the reversal of the polar field is comparatively easier and there is no overlap between two cycles. However, \cite{GBKKK23} have shown that there is a significant variation in the reversal timings ($T_r$) of different cycles.
It has also been observed that sun's polar field reversed multiple times in some cycles \citep{MFS83, Mord22}.  We also found this kind of multi-reversal in our simulations due to a large scatter in BMR tilt. In those cases, we take the time of the first reversal as our reference time.
From $T_r$ we take the simulated polar field data for the next three years (the green vertical line in panel (b) indicates the time of prediction, $T_p = T_r+3$ years) and calculate the rise rate of the polar field in those three years between $T_r$ and $T_p$.
The polar fields do not rise in a uniform manner throughout those three years, but show rapid variations in the initial years. To tackle this issue and to get an average rate of rise of the polar fields, we divide these three years into several overlapping segments and then calculate the rise rate in each of these segments. Finally, the mean rise rate is taken as the ultimate rise rate of the polar fields.
In the next step, we find the peak value of the polar field at the end of each cycle and calculate the evolution of the toroidal field to be generated in the next cycle by integrating \Eq{tor}. For the amplitude of the next cycle, we take the peak of the calculated toroidal field.
Finally, we measure the correlations between the peak strength of the polar field and the peak of the next cycle's toroidal field with the rise rate of the polar field.  For different cases of our study we analyze the impact of fluctuations in the various aforesaid parameters on the correlation between these quantities.

\section{Results and Discussions}
Before we start discussing the results from the simulations, we would like to mention the features of the solar cycle that can potentially induce the variation in the polar field build-up and subsequently in the polar field rise rate. Firstly, the amplitude of the solar cycles has been observed to vary significantly which leads to variation in the rate of BMR emergence and in turn, in the eventual build-up of the polar field. Secondly, the meridional circulation is a key driver behind the transport of the residual field of the disintegrating BMRs toward the poles. Hence, any cycle-to-cycle variation in the amplitude of meridional circulation can impact the growth of the polar field. Lastly, and most importantly, the large scatter in the BMR tilts around Joy's law and the variation in the presence of `anomalous' BMRs can produce significant variation in the aforesaid quantities.

Here, in this section, we will discuss our analysis from the different cases of the simulations using different realizations of the essential parameters of BMR properties and of the Model that can impact the polar field build-up as mentioned above. A list of the six major Cases of the study is given below:
\begin{itemize}
    \item[I)] Variation in cycle amplitudes with BMR tilts strictly following Joy's law.
    \item[II)] Same as Case I, but with varying meridional flow speed in each cycle, i.e., variation in cycle amplitude and in meridional flow speed with BMR tilts strictly following Joy's law.
    \item[III)] Variation in the BMR tilt scatter with anti-Hale BMRs present in all the phases of the cycles.
    \item[IV)] Same as Case III, but with all anti-Hale BMRs in the rising phases of the cycles.
    \item[V)] Same as Case III, but with all anti-Hale BMRs in the declining phases of the cycles.
    \item[VI)] Same as Case III, but with the BMR tilt scatter dependent on the area of the BMRs.
\end{itemize}
In all of the five Cases, the amplitudes of the cycles vary within the range of 30 to 90 in terms of monthly BMR number. Here we mention that in the Case IV and Case V, we include all the 3--7\% anti-Hale BMRs within the rising and the decline phases of the cycles. Hence, the temporal density of anti-Hale BMRs is significantly higher in the last two cases than the Case III (see \Fig{bfly_ah}) or observations. This is probably not a reality for the actual solar cycles where the anti-Hale BMRs may not show any preference in their presence to the phases of the cycles. However, we create this hypothetical worst-case scenario to test the robustness of the above-mentioned correlation under the impact of a significantly increased temporal density of anti-Hale BMRs in a certain phase of the cycle.

\begin{figure*}
    
    \includegraphics[scale=0.5]{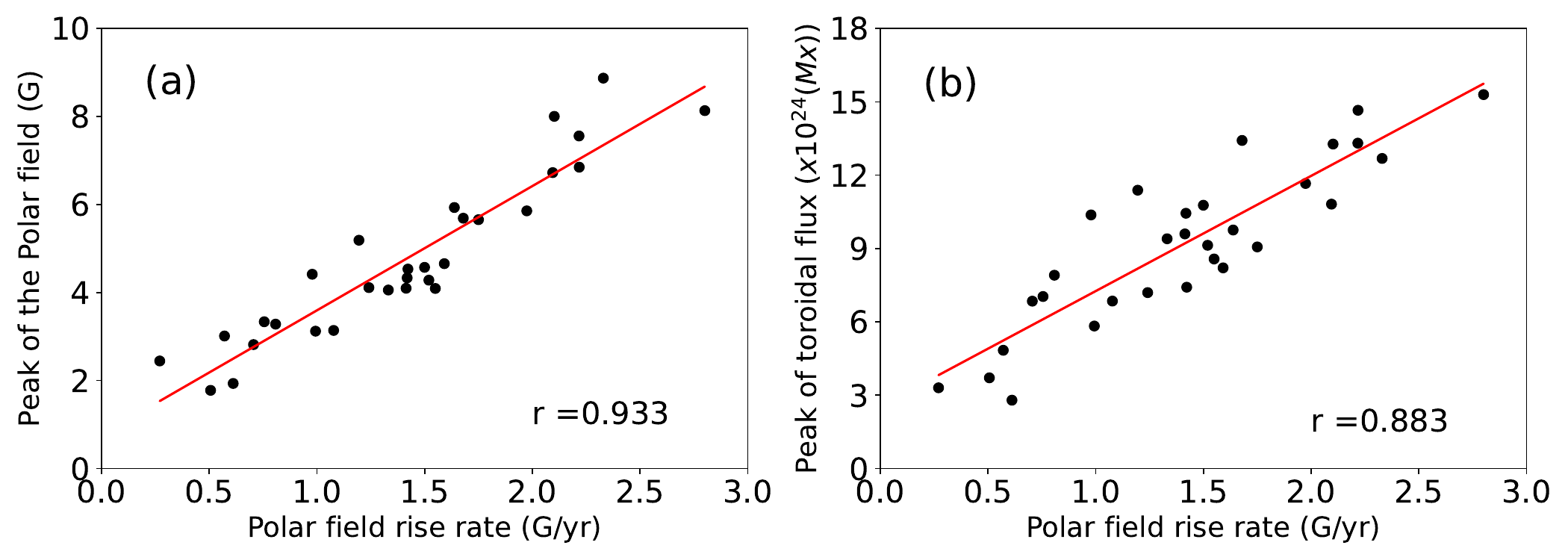}
    \caption{Plot between the rise rate of polar field and the peak of the polar field (a) and the peak of the next cycle's toroidal field to be produced by the polar field (b). These results are obtained from simulations with variations in cycle amplitudes  (Case I).  }
    \label{amp_vary}
\end{figure*}

\begin{figure*}
    
    \includegraphics[scale=0.5]{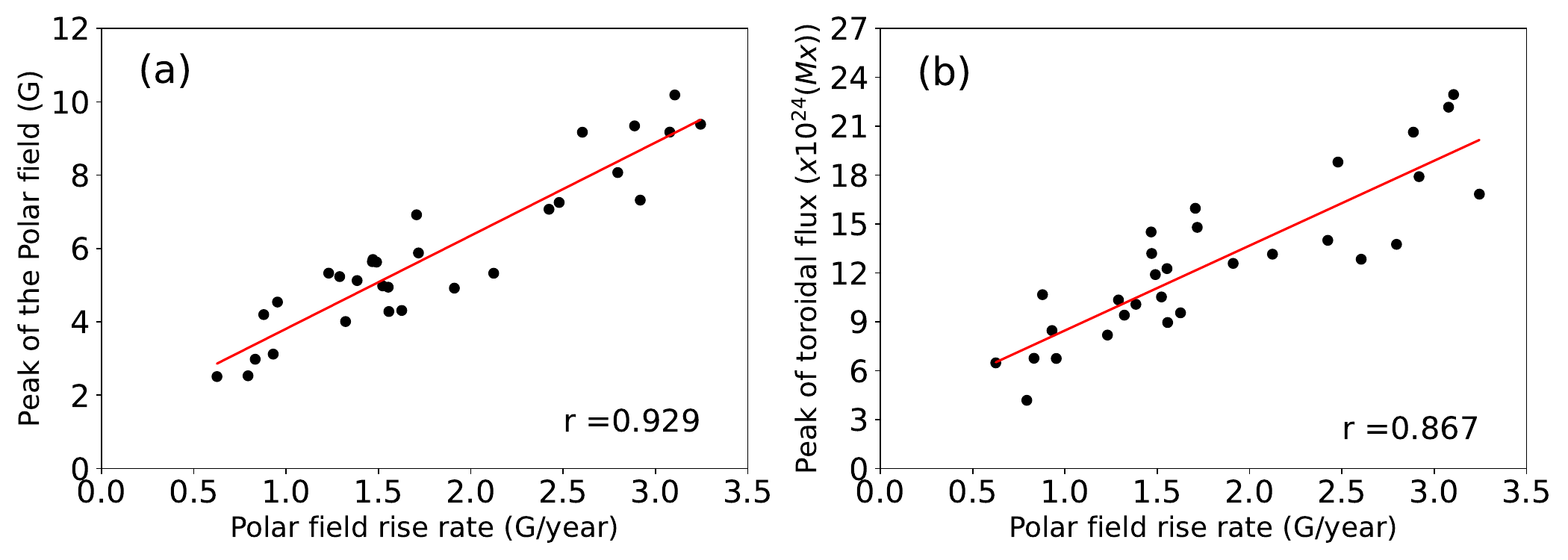}
    \caption{
    Same as \Fig{amp_vary} but for case II, i.e. with variation in both the amplitude and in the meridional flow speed of the cycles.}
    \label{merid_vary}
\end{figure*}

In \Fig{amp_vary}, we present the results obtained from the simulations for Case~I. In this case, the tilts of the BMRs are obtained from Joy's law (the black straight line in panel (c) of \Fig{cycle_tilt})  i.e., there are no anomalous (anti-Joy or anti-Hale) BMRs in these cycles and the meridional flow speed is the same for all the cycles. However, the amplitudes of the cycles are different from each other which makes the spatiotemporal profile of the BMRs to vary from one cycle to the other. This variation in the BMR profile leads to the variation in the build-up of the polar field. It can be seen that the peak of the polar field at the end of each cycle and the strength of the produced toroidal field in the following cycle both are very highly correlated with the rise rate of the polar field build-up. A high value of correlation between these quantities infers a better predictability of the strength of a  solar cycle from the rise rate of the polar field. In the following four cases, we will include additional stochastic fluctuations in different parameters involved in the model and BMR profiles to examine their impact on this correlation.

Now, we discuss the results from Case~II. The meridional flow speed ($v_0$) for the cycles are randomly assigned from a uniform distribution between $10 \le v_0 \le 30$ m/s. In this case also, the tilts of the BMRs strictly follow Joy's law. 
The scatter plots obtained from the simulations are presented in  \Fig{merid_vary}. We get a high value of linear Pearson correlation as mentioned in the individual panels of the figure. 
This result suggests that the variation in meridional circulation 
does not decrease the correlation between the rise rate and the peaks of the aforesaid quantities, hence it does not have any adverse impact on the predictability of the solar cycles using the rise rate of polar field.

Before we start the discussion of the rest of the three cases, we present the respective butterfly diagrams presenting the spatiotemporal profiles of the BMRs in each of these cases in \Fig{bfly_ah}. The panels (a), (b), and (c) represent the typical butterfly diagrams of cases III, IV, and V, respectively. The color bar in the butterfly diagrams shows the value of BMR tilts, where the blue colored spots are the anti-Joy BMRs with $-90^\circ<\gamma<0^\circ$, and the slightly bigger black dots represent the anti-Hale BMRs with $-180^\circ<\gamma<-90^\circ$. In all these three cases, we take the different realization of the tilt properties and the different percentages of anomalous regions in different cycles. Typically, 
the percentage of the anti-Joy regions varies within a range of roughly $10-30\%$ of the total number of BMRs, on the other hand, we choose the percentage of anti-Hale BMRs within $3-7\%$, consistent with observations \citep{McClintock+Norton+Li14}.
The meridional flow speed is same for all the cycles in all three cases with $v_0 = 22$ m/s.

\begin{figure*}
    
    \includegraphics[scale=0.2]{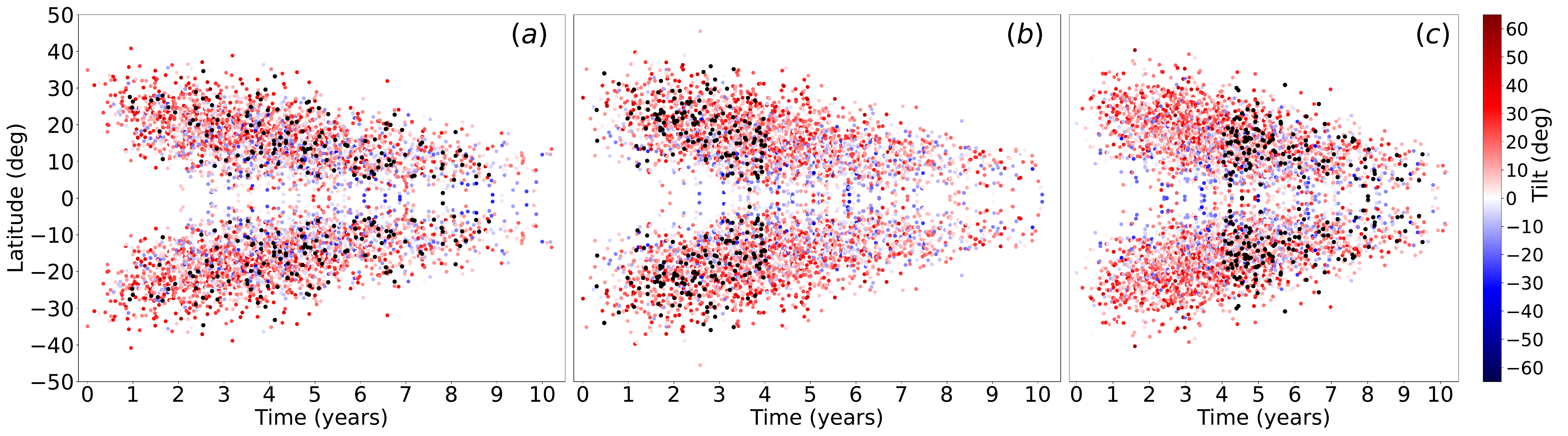}
    \caption{Butterfly diagrams with the different types BMRs represented by the color schemes depending on their tilt. The red BMRs are the regular ones with a positive tilt, whereas, the blue colored dots represent the anti-Joy BMRs with a small negative tilt and the bigger black dots are the anti-Hale BMRs.  The panels (a), (b), and (c) represent the cases III, IV, and V, respectively.   Note that the colorbar is saturated beyond $\pm60^\circ$ and thus all anti-Hale BMRs appear as black dots.
    }
    \label{bfly_ah}
\end{figure*}

\begin{figure*}
    
    \includegraphics[scale=0.5]{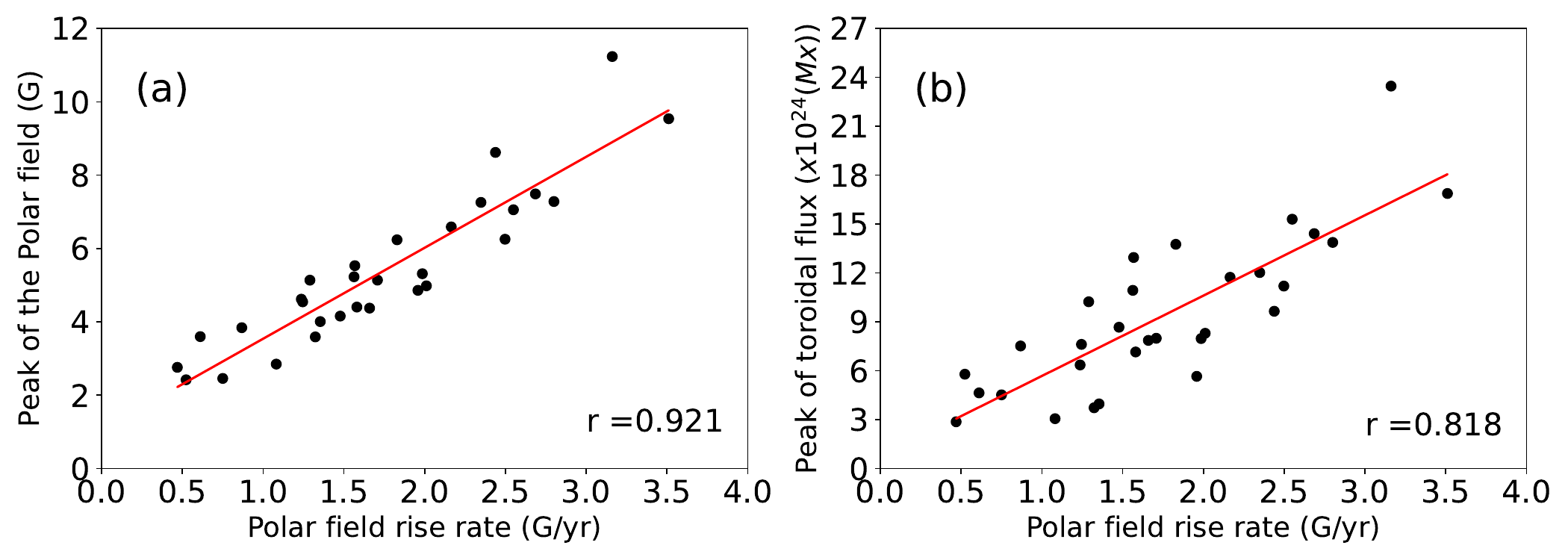}
    \caption{Same as \Fig{amp_vary} but for case III, i.e. with anti-Hale BMRs deposited throughout all the phases of the cycles}
    \label{ah_all}
\end{figure*}

Next, we discuss the results from Case~III where we randomly deposit the anti-Hale BMRs throughout all the phases of the cycles. The scatter plots for this case is shown in \Fig{ah_all}. The value of Pearson correlation coefficients of the polar field rise rate with the peak of the polar field and the peak of the toroidal field is high in this case as well. This result implies that the predictability of the following solar cycle strength does not get affected much even in the presence of significant tilt scatter and the anti-Hale BMRs. 

\begin{figure*}
    
    \includegraphics[scale=0.5]{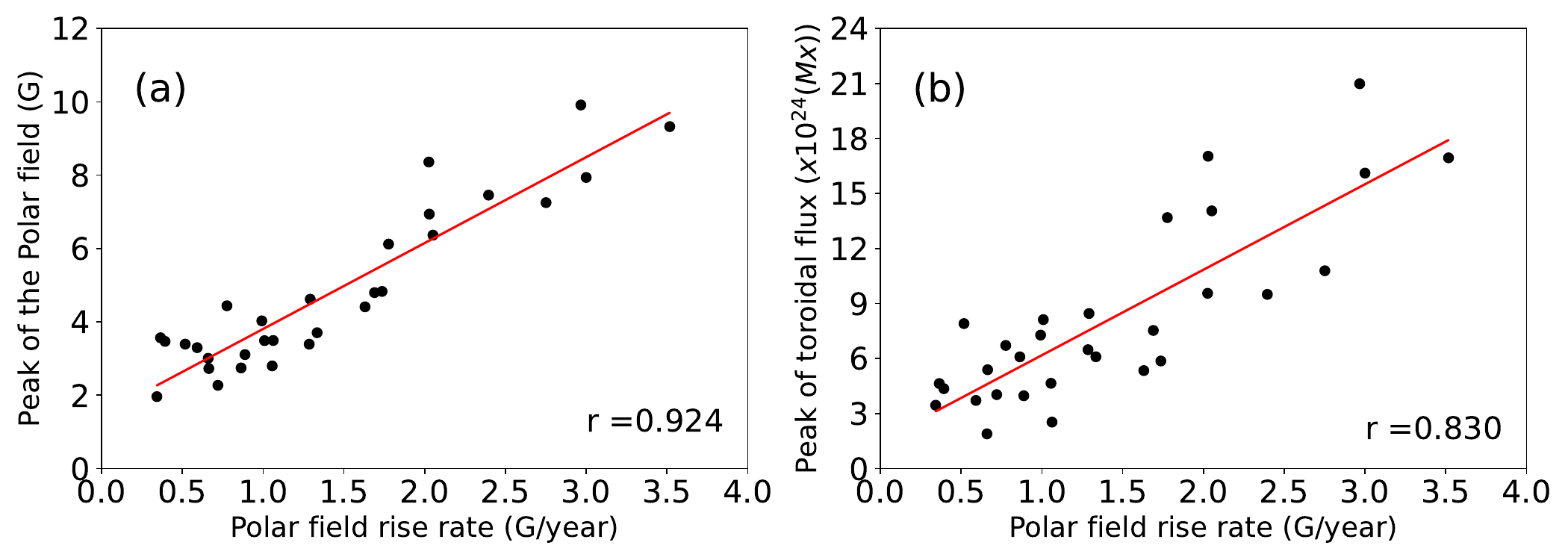}
    \caption{Same as \Fig{amp_vary} but for case IV, i.e. with anti-Hale BMRs deposited only in the rising phases of the cycles}
    \label{ah_rise}
\end{figure*}

\begin{figure*}
    
    \includegraphics[scale=0.5]{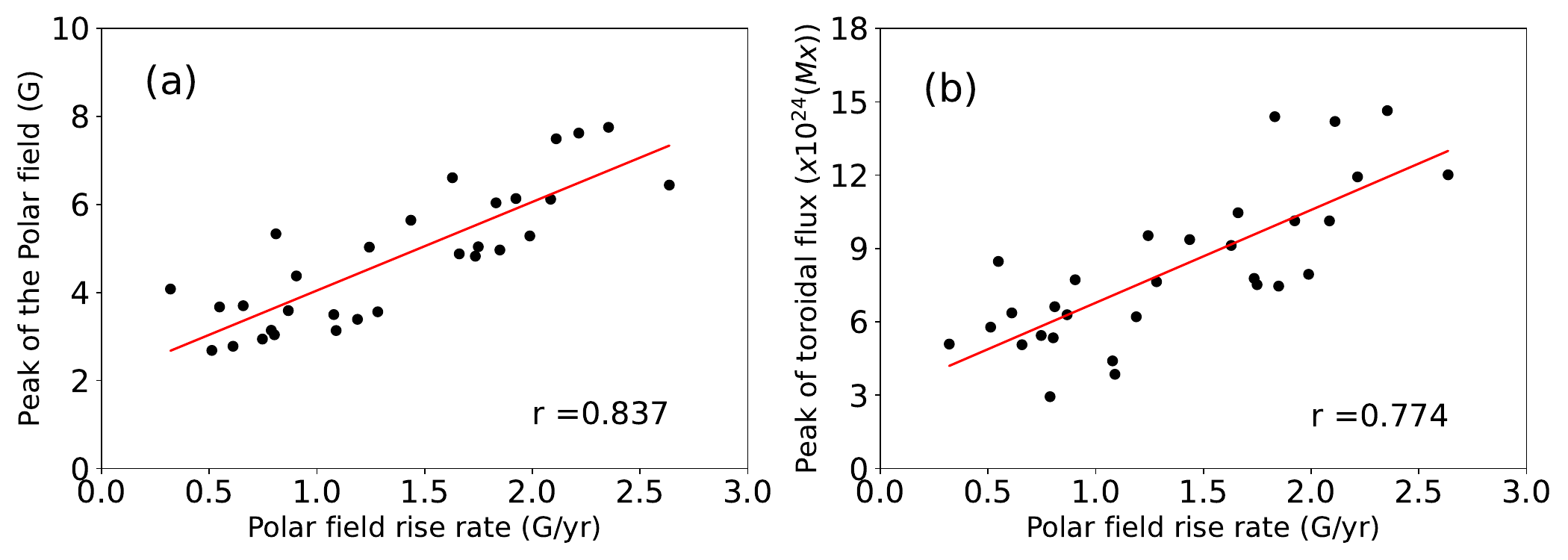}
    \caption{Same as \Fig{amp_vary} but for case V, i.e. with `all anti-Hale BMRs' deposited only in the declining phases of the cycles}
    \label{ah_dec}
\end{figure*}

In \Fig{ah_rise} we show a similar plot but this time, all the anti-Hale BMRs are deposited only in the rising phases of the cycles (case IV). Again, in this case also, the correlation coefficients between the plotted quantities are high and similar in values to the previous two cases. 
Hence, having a significant amount of anomalous BMRs in the beginning phases of the cycles does not hamper the predictability of the polar field at the cycle minima or the next cycle strength from the polar field rise rate.
This is easy to understand because the polar field generated in the rising phase of the cycle is mostly used to reverse the old polarity field and thus the build-up of the polar field after its reversal is undisturbed.

\begin{figure}
    
    \includegraphics[scale=0.43]{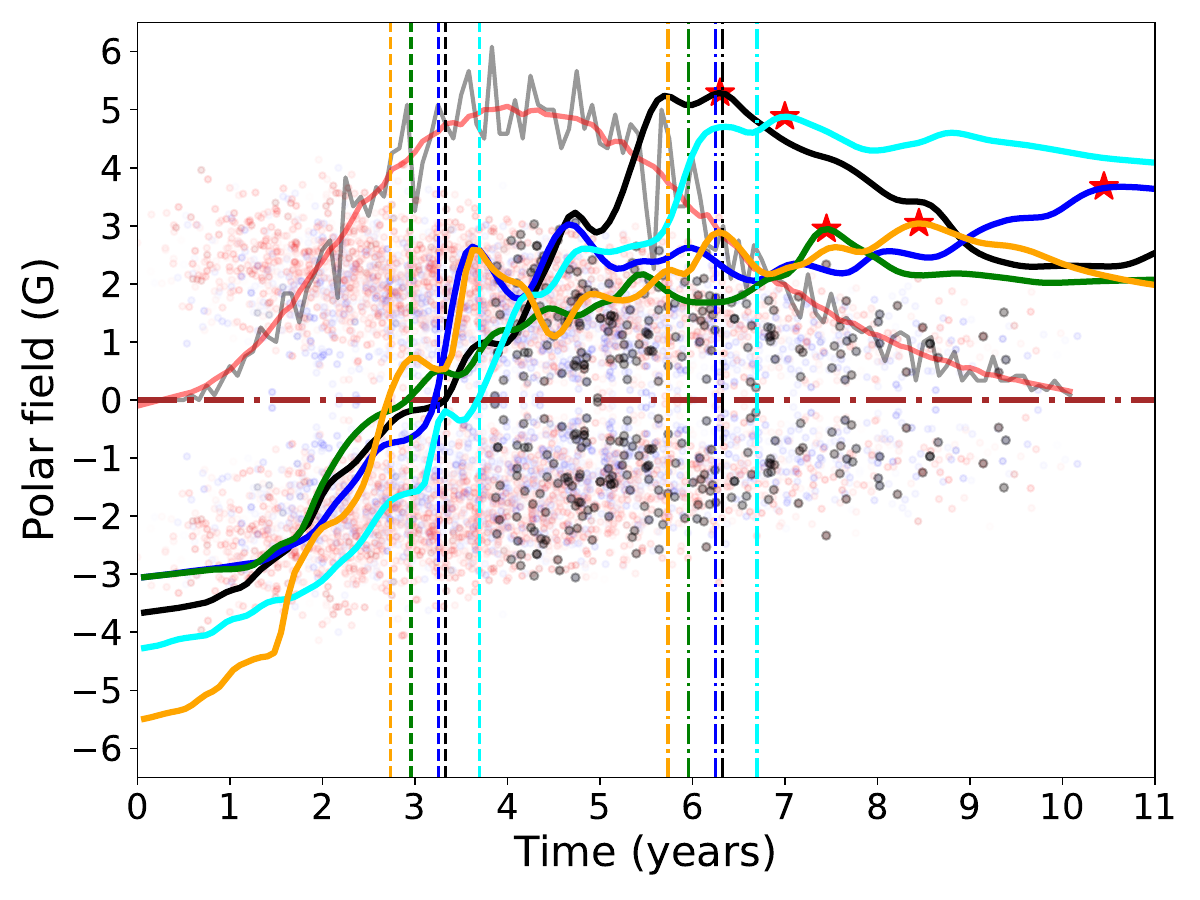}
    \caption{The evolution of northern polar fields for typical five cycles belonging to Case~V. Vertical dashed and dashed-dot lines respectively represent the $T_r$ and $T_p$ of the individual cycles. The typical variation of the BMR number (similar to panel (a) of \Fig{cycle_tilt}) is shown in the background for the temporal reference regarding the phase of the solar cycles. The butterfly diagram with the tilt information (similar to panel (c) of \Fig{bfly_ah}) has also been presented in the background. The red asterisk symbols represent the peaks of the polar fields. }
    \label{pf_dec}
\end{figure}

Next, the scatter plots for case V, having all the anti-Hale BMRs deposited at the declining phases of the cycles are shown in \Fig{ah_dec}. 
In this case, the trends between the plotted quantities are similar to the previous cases, however,  a significant drop in the values of the correlation coefficients is observed. 
This is somewhat expected, as there is a very high concentration of anomalous BMRs during the ending phases of the cycles which significantly diverts the trajectory of the polar field, causing the actual peak to be far from the expected value. 
The typical evolution of the polar fields for five cycles belonging to Case V is presented in the \Fig{pf_dec}. The vertical dashed lines show the reversal time ($T_r$) and the vertical dashed-dot lines show the prediction time ($T_p$).
In the background, the profile of the solar cycle is shown for reference to the different phases of the cycle along with the typical butterfly diagram. 
From the evolution of the polar fields, it can be easily inferred that the phase before the reversal exhibits a nearly undisturbed evolution as in this phase the presence of anomalous regions is not significant. 
On the other hand, just after the reversal, due to the presence of the anti-Hale regions, the growth of the polar field is severely disturbed.
As a result, the correlation between the plotted quantities gets worsened.
However, the values of the correlation coefficients suggest that a decent estimate of the strengths of the polar field and that of the following cycle can be obtained from the rise rate even in this case as well, but with reasonably wider error bars. 

It has been observed that the tilt scatter or the standard deviation ($\sigma$) of the tilt distribution shows a significant and systematic dependence with the area of the BMRs. To check whether this dependence impacts the predictability of the solar cycle, we perform the simulations for the Case (VI) where we have re-investigated Case III with the area-dependent $\sigma$ of the tilt distribution which is obtained from the formula given by \cite{JCS14}. In this case, from the simulation of 30 cycles, we have obtained strong correlation coefficients of 0.903 and 0.851 between the polar field rise rate and the polar field amplitude and the following cycle amplitude respectively. This result emphasizes that there is not much impact of the area-dependent tilt scatter on the correlation between these quantities. The $p$ values of all these correlations have not been presented as their values are much smaller ($<10^{-7}$) than the significance level.

\section{Conclusion}

In conclusion, we find that the strong correlation of the polar field rise rate a few years after the cycle maximum with the peak of the polar field (at the cycle minimum) and the strength of the following sunspot cycle, as first shown in \citet{KBK22}, is a robust feature of the solar cycle. 
Our extensive simulations using SFT model suggest that this relation between the aforesaid quantities can be utilized to reliably forecast the strength of the upcoming solar cycle much earlier than the cycle minimum. 
For our study, we have taken data from only the first three years from the first reversal of the polar field  ($T_r$ to $T_p$) which happens during the cycle maximum.
It is to be noted that, even in the cases of multi-reversals in the polar field, we get the high correlation between the above quantities by taking the first reversal time as $T_r$.
We find that cycle-to-cycle variation of meridional flow speed does not have any impact on the predictability of the solar cycle using this method.
We get a very high correlation between the peak of the following cycle with the rise rate of the polar field even in the presence of anomalous BMRs throughout the cycles consistent with observation.
However, a high concentration of anomalous BMRs present during the declining phase of the cycles hampers this correlation to some extent, but still, the correlation is strong enough to provide a reasonable forecast of the upcoming cycle's strength.
 
\section{Acknowledgements}
The authors are grateful to Dr. Robert Cameron for kindly providing the SFT code used in this study and for providing important insights regarding the results during several discussions. The authors also thank Prof. Kristof Petrovay for fruitful discussions that benefitted this project. The authors are thankful to the anonymous referee for carefully reviewing the manuscript and for providing insightful suggestions to improve the quality of the paper.    A. B. acknowledge financial support from the University Grants Commission, Govt. of India. A.B. and B.B.K. acknowledge the financial supports from the ISRO/RESPOND program (project No. ISRO/RES/2/430/19-20) and Ramanujan Fellowship (project no SB/S2/RJN-017/2018).
This study benefitted from financial support from the International Space Science Institute (ISSI Team 474).

\section{Data Availability}
In this work, we have performed the simulations using the SFT code as mentioned in \citet{Bau04}. The code for the synthetic BMR profile and the data from the SFT simulations along with the codes for the data analysis can be shared upon a reasonable request.



\bibliographystyle{mnras}
\bibliography{paper} 




\appendix


\bsp	
\label{lastpage}
\end{document}